\documentclass[twocolumn,superscriptaddress,floatfix,preprintnumbers,nofootinbib,nobibnotes,amsmath,amssymb,aps]{revtex4-1}

\usepackage{graphicx}
\usepackage{dcolumn}
\usepackage{bm}
\usepackage{hyperref}
\usepackage[mathlines]{lineno}
\usepackage{color}
\usepackage{xcolor}
\usepackage{verbatim}
\usepackage{soul}

\definecolor{brown}{rgb}{.7,.35,.1}

\begin{document}

\preprint{FERMILAB-PUB-25-0416-T}

\title{Gravitational Photon Polarization Twist to Probe the Early Universe and the Galactic Center}

\author{Rodolfo Capdevilla}
\email{rcapdevi@fnal.gov}
\affiliation{Theoretical Physics Division, Fermi National Accelerator Laboratory, Batavia, IL 60510, USA}

\date{\today}

\begin{abstract}
It is known that gravitational wave backgrounds (GWBs) change the polarization state of photons. This letter explores the possibility of using this effect to detect GWBs. The proposed experiment features a vertically polarized laser pulse traveling through a GWB before reaching a birefringent material which separates photons by polarization. Photons that emerge the birefringent material horizontally polarized are counted as signal photons. A space-based setup of a million km (comparable to the LISA mission) with a millisecond laser pulse is capable to detect the hypothetical pulsars responsible for the excess of gamma rays from the galactic center. Varying the duration of the pulse can reveal a variety of GWBs from the early Universe as well.
\end{abstract}

\maketitle

{\bf Introduction} - The discovery of gravitational waves (GWs) has confirmed that gravitational perturbations propagate as predicted by General Relativity. Interferometers LIGO and Virgo detect GWs in the 1-$10^3$ Hz frequency range~\cite{LIGOScientific:2016aoc} originated from stellar black hole (BH) mergers. Future missions like LISA~\cite{LISACosmologyWorkingGroup:2022jok} will measure GWs down to frequencies of $10^{-4}$ Hz. In addition to BHs, pulsars also generate GWs within the frequency range of current experiments~\cite{Agarwal:2022lvk}. Indeed, if an unresolved population of pulsars were responsible for the Galactic Center gamma-ray excess (GCE)~\cite{Goodenough:2009gk,Abazajian:2012pn,Eckner:2017oul,Ploeg:2021cqk}, they would produce a secondary signal; a flux of GWs with a peak amplitude at around 1 kHz~\cite{Calore:2018sbp,Miller:2023qph,Carleo:2023qxu,Bartel:2024jjj}. However, even next-generation interferometers will likely not have the sensitivity to measure this signal~\cite{Bartel:2024jjj}. Since its discovery, it has been suggested that the GCE is the product of dark matter (DM) annihilation~\cite{Hooper:2011ti,Abazajian:2012pn,Gordon:2013vta,Daylan:2014rsa}. Hence, while detecting a GW signal from the galactic center would confirm pulsars as the GCE source, a null result would further strengthen the DM hypothesis as the leading explanation.

Besides BHs and pulsars, plasmas are other compelling sources of GWs~\cite{Ghiglieri:2015nfa,Giovannini:2019oii,Ghiglieri:2020mhm,Cabral:2016klm,Garg:2022wdm,Garg:2023yaw}, e.g., the Sun~\cite{Garcia-Cely:2024ujr} and neutron stars~\cite{McDonald:2024nxj}. Similarly, when the Universe was a hot plasma, reaching temperatures of at least a few MeV, it likely left a relic Cosmic Gravitational Wave Background (CGB). The CGB has a spectrum today that peaks at 10-100 GHz~\cite{Ghiglieri:2015nfa}. Known high-frequency GW (HFGW) sources, like the CGB and stars, have extremely small strains which makes them very challenging targets. One approach to detect them involves leveraging the Gertsenshtein effect~\cite{Gertsenshtein:1961xxx}, where gravitons convert into photons under the influence of a magnetic field. This has been investigated in the context of axion DM experiments over recent years~\cite{Arvanitaki:2012cn,Aggarwal:2020umq,Carney:2024zzk,Goryachev:2014yra,Kahn:2023mrj,Berlin:2023grv,Schnabel:2024hem,Ballantini:2003nt,Quach:2016uxd,Ito:2019wcb,Domcke:2022rgu,Bringmann:2023gba,Domcke:2023bat,Berlin:2021txa,Navarro:2023eii,Ahn:2023mrg,Gatti:2024mde,Fischer:2024msc,Domcke:2024eti,Capdevilla:2024cby,Ejlli:2019bqj,Capdevilla:2025omb,Krokotsch:2025iyi,Pappas:2025zld,Fischer:2025mpz}. So far, these experiments lack the sensitivity to detect known HFGW sources. Since these experiments often push quantum systems to their limits, increasing the sensitivity as needed for HFGWs may be technologically unfeasible, suggesting the need for novel detection mechanisms. This is the spirit of this letter.

\begin{figure}[t]
\centering
\includegraphics[width=0.485\textwidth]{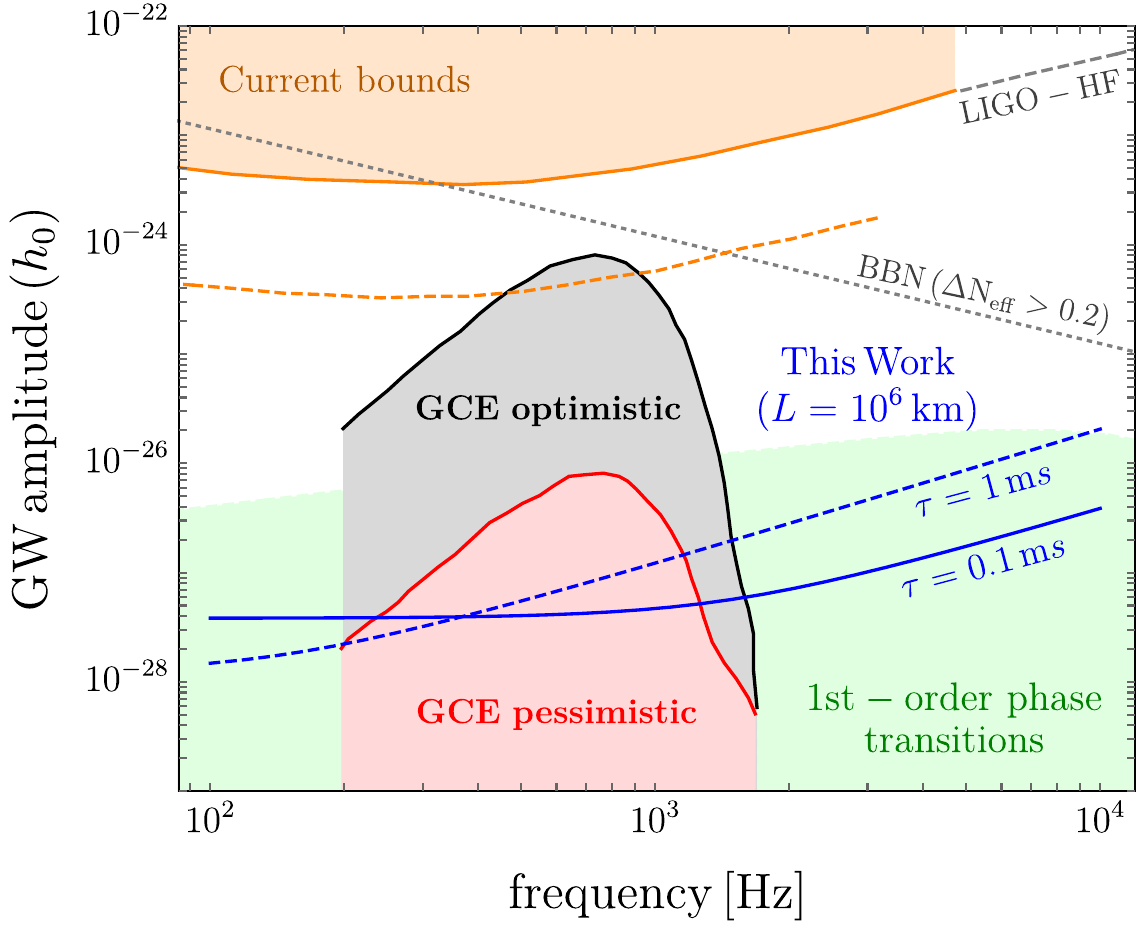}
\caption{The hypothetical GCE pulsars emit the black (red) spectrum in the optimistic (pessimistic) scenario described in~\cite{Bartel:2024jjj}. Current (solid) and projected (dashed) sensitivity from interferometers are in orange. First-order phase transitions in the early Universe can produce a signal within the green contour. The experiment proposed in this letter can potentially reach down to the blue curves. The baseline $L$ is the distance between the laser source and the birefringent material, and $\tau$ is the pulse duration. The blue lines assume a 800 nm and 30 kW laser as a benchmark. Even reducing the laser power down to 2 W (LISA-like) the proposed experiment is able to detect the GCE pulsars.}
\label{fig:sensitivity}
\end{figure}

\begin{figure*}[t]
\centering
\includegraphics[width=0.85\textwidth]{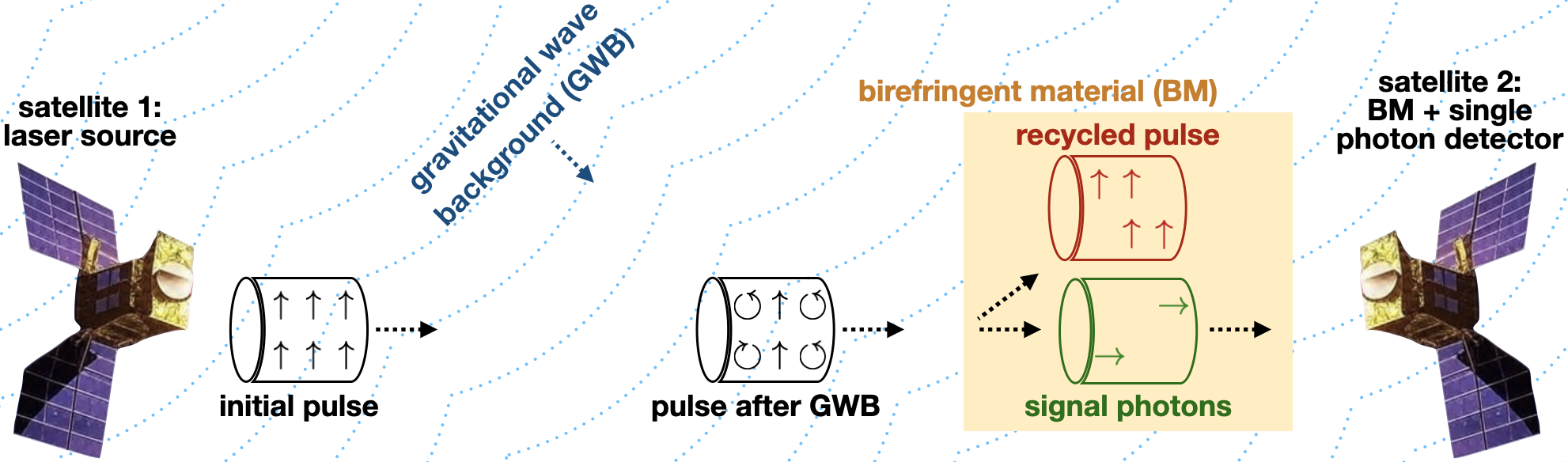}
\caption{Sketch of the proposed experiment: A vertically polarized pulse (satellite 1) interacts with a GWB and some photons in the pulse transition to circular polarization. Half of the photons that transitioned pass through a birefringent material (satellite 2) as horizontally polarized and are read as signal photons by a single photon detector. If possible, the recycled pulse could bounce between satellites and increase the number of signal photons.}
\label{fig:setup}
\end{figure*}

In the proposed experiment, a stream of linearly polarized photons confined within a laser pulse scatter off a GWB. A fraction of the photons transition into circular polarization states~\cite{IACOPINI1979140,mnras,AMCruise_2000}. This effect is what this letter refers to as {\it Gravitational Photon Polarization Twist}. As it is demonstrated below, the effect depends on the baseline $L$; the longer the pulse is immersed in the GWB, the more photons {\it twist} their polarization. Based on conservative estimates, a setup with a baseline comparable to that of the LISA mission will be able to detect the GWs signal from the galactic center, in case pulsars are the GCE source. This can be seen in Fig.~\ref{fig:sensitivity} as the blue lines reach down to the black and red curves. The later correspond to the GW spectra by the hypothetical pulsars; the red curve~\cite{Miller:2023qph} assumes the most realistic determination of the luminosity function that determines the number of millisecond pulsars required to explain the GCE, whereas the black curve~\cite{Bartel:2024jjj} is a more optimistic estimate within astrophysical uncertainties.

Beyond the aforementioned sources within the Standard Model, a range of new physics phenomena can generate GWs. These include primordial BHs~\cite{Dolgov:2011cq,Franciolini:2022htd}, exotic dark sectors~\cite{Giudice:2016zpa,Landini:2025jgj}, primordial magnetic fields~\cite{Fujita:2020rdx}, cosmic strings~\cite{Servant:2023tua,Fu:2024rsm}, modified gravity~\cite{Delgado:2025ext}, inflation~\cite{Garcia:2024zir,Saha:2024lil}, and 1st-order phase transitions~\cite{Caprini:2010xv,Schwaller:2015tja,Kobakhidze:2017mru,Roshan:2024qnv,Feng:2024pab}. The latter can reveal information about the origin of the matter anti-matter asymmetry. The amount of GWs that these Cosmological sources (from {\it early} Universe) can produce is constrained by observations such as the cosmic microwave background (CMB) and primordial nucleosynthesis~\cite{Planck:2018vyg,Planck:2018jri,Kawasaki:1999na,Kawasaki:2000en,Hannestad:2004px}. Astrophysical sources (from {\it late} Universe) evade these constraints. One example is BH superradiance~\cite{Arvanitaki:2009fg,Arvanitaki:2010sy,Yoshino:2013ofa}, where rotating BHs emit bosonic particles (e.g., axions) which form a {\it gravitational atom} and emit GWs. The experiment proposed in this letter has the potential to discover new physics from any of these sources, as demonstrated by the blue lines extending through the green contours in Fig.~\ref{fig:sensitivity}. The possibility of not only resolving the mistery behind the GCE, but also discovering new physics, underscores the rich scientific case for the proposed experiment.

In the following a few discussions are in order; first, the proposed experimental setup, second, the estimated sensitivity of the experiment under reasonable assumptions, and finally, the details of the calculation are shown right before summary and conclusions.\\

{\bf Signal and Experimental Setup} - After a vertically polarized laser pulse overlaps with a GW, the number of photons that transition to states of circular polarization, hereafter called {\it signal photons}, is given by
\begin{equation}
N_s=N_\gamma h_0^2\omega_L^2s_{\theta}^4\frac{\pi}{2}\tau L\,\,\eta,
\label{eq:Ns}
\end{equation}
where $N_\gamma$ is the number of initial photons in the pulse, $h_0$ the GW strain, $\omega_L$ the laser frequency, $\theta$ the angle between the propagation of the laser and the GW, $\tau$ the pulse duration, and $\eta$ is an order-one parameter as long as the length of the pulse is of the order of the GW wavelength (details provided below). The following choice of parameters gives $N_s\sim1$ (averaging over all GW angles)
\begin{equation}
N_s \sim 1 \times \left( \frac{N_\gamma}{10^{20}} \right) \left( \frac{h_0}{10^{-23}} \right)^2 \left( \frac{\tau}{1 \text{m}s} \right) \left( \frac{L}{10^6 \text{km}} \right).
\label{eq:NsBench}
\end{equation}
For example, for a $\omega_L = 1.55 \, \text{eV}$ ($800 \, \text{nm}$ wavelength) laser to contain $N_\gamma = 10^{20}$ photons, it requires 30 J. Now, a millisecond pulse of this energy implies 30 kW of power. For comparisons, the LISA mission will use a continuous laser with 2 W of power, whereas LUXE~\cite{Abramowicz:2021zja,LUXE:2023crk} will use a 3 PW source~\cite{fede} delivering 75 J per each 25 fs pulse. As discussed below, even a 2 W laser can produce enough signal events to discover the GCE pulsars.

Provided the long distances considered to achieve $\mathcal{O}(1)$ signal photons, it is clear that the proposed experiment needs to take place in space. To estimate the background one needs to compute the mean free path of the pulse. The number density of deep space is $\sim$10 charged particles (protons/electrons) per cm$^3$~\cite{EVIATAR1970321,Wilson2018,Mann2010}. Charged particles interact with the pulse via Thomson scattering, with cross-section $\sigma = 8\pi \alpha^2 / 3 m^2 \sim 10^{-24} \, \text{cm}^2$ \cite{jackson1998classical2}, where $\alpha$ is the fine-structure constant and $m$ is the particle mass. For electrons this yields a mean free path of $\ell_{\rm MFP} = 1 / (\sigma n) \sim 10^{18} \, \text{km}$, vastly exceeding the $10^6$ km benchmark. As background contributions from space particles are negligible, the dominant noise source is the dark count rate (DCR) of the single-photon detector (SPD) that will read the signal photons. DCRs can be as low as $10^{?4}$ Hz for advanced quantum sensors \cite{Verma:2020gso,Wollman:17,PhysRevD.88.035020}.

To measure the signal photons, this letter proposes the experiment in Fig.~\ref{fig:setup}. Birefringent materials have polarization dependent index of refraction~\cite{GUOQING1998517,GHOSH199995,Zhang2022}, which can be leveraged to isolate the signal photons. As depicted in Fig.~\ref{fig:setup}, as the laser pulse enters the BM, the vertical and horizontal polarizations split. Because the signal photons are circularly polarized, half of them (on average) will follow the path towards the SPD. The Fig.~\ref{fig:setup} setup can serve two purposes; to collect the signal photons, and to {\it recycle} the pulse and repeat the experiment with mirrors that would make the pulse bounce between satellites and pass through the BM multiple times.

When the pulse passes through the BM a number of $\ell$ laps, the number of photons in it is given by $N_\gamma^\ell=\epsilon_{\rm BM}^\ell N_\gamma$, where $\epsilon_{\rm BM}$ is the transmission efficiency of the BM. The total signal photons are the sum over all laps
\begin{equation}
N_s^{\rm tot}=\sum_{\ell=0}^{\ell_{\rm tot}}{N_s \over N_\gamma} N_\gamma^\ell=\sum_{\ell=0}^{\ell_{\rm tot}}\epsilon_{\rm BM}^\ell N_s,\label{eq:Nstot}
\end{equation}
where $N_s$ is given by Eq.~(\ref{eq:Ns}). A reasonable value for $\epsilon_{\rm BM}$ today can be found in~\cite{Sakakura2020}, where it is shown that induce birefringence in silica glass can  reach transmission efficiencies of 99\% over a wide range of wavelengths (400-1200 nm). However, more recent work~\cite{Lei2023} has demonstrated that this value can be improved, potentially up to 99.9\%~\cite{sakakura}. A benchmark of $\epsilon_{\rm BM}=0.999$ could increase $N_s$ in Eq.~(\ref{eq:NsBench}) by a factor of a $10^3$.\\

{\bf Estimated Sensitivity} - The sensitivity of the Fig.~\ref{fig:setup} setup can be estimated by setting a signal-to-noise (SNR) larger than five over an observation time $t_{\rm obs}$, where
\begin{equation}
{\rm SNR} = \sqrt{t_{\rm obs}}{S \over \sqrt{B}}>5,
\label{eq:snr}
\end{equation}
and $S/B$ are the signal/background rates. One Hz is a reasonable rate for a high-power laser pulse~\cite{LUXE:2023crk}. Replacing $S=N_{s}^{\rm tot}\,{\rm Hz}$ and $B=10^{-4}\,{\rm Hz}$ (the DCR) into Eq.~(\ref{eq:snr}) one can estimate the sensitivity of the Fig.~\ref{fig:setup} setup, by computing the minimal strain $h_0$ that could provide a signal above the noise, given by
\begin{equation}
h_0^2 > {{\rm SNR} \over 10^{3}N_\gamma\omega_L^2s_{\theta}^4(\pi/2)\tau L}  \sqrt{{10^{-4} \over t_{\rm obs}\,{\rm Hz}}}.\label{eq:h_lab}
\end{equation}

For one year of observation and similar parameters to Eq.~(\ref{eq:NsBench}) one can get ${\rm SNR}>5$ for strains
\begin{equation}
h_0 > 5\times10^{-28}.
\label{eq:h_space}
\end{equation}

This estimate is conservative, assuming a GWB source with a single frequency and strain. In practice, the GWB strain exhibits a spectral profile, as shown e.g. by the GCE pulsars in Fig.\ref{fig:sensitivity}. Integrating over this spectrum enhances the sensitivity of the experiment beyond the naive prediction of Eq~(\ref{eq:h_space}). Now, using the collective strain of the source as a proxy for $h_0$, the pessimistic GCE curve in Fig.\ref{fig:sensitivity} yields an integrated strain of $\sim$$10^{?26}$. As the initial photon number $N_\gamma$ scales linearly with laser power and sensitivity varies as $\sqrt{N_\gamma}$, even if instead of a 30 kW one uses a 2 W laser, which reduces the signal photons by four orders of magnitude, the proposed experiment still reaches the sensitivity to detect the GCE pulsars. Alternatively, if recycling the pulse is technologically unfeasible, a 3 kW laser suffices for the same discovery.

Following the same conservative approach, the projected sensitivity of the Fig.~\ref{fig:setup} setup appears in blue in Fig.~\ref{fig:sensitivity}. Current bounds from interferometers are marked by the orange shadow, while future projections are given by the dashed orange line. The dashed gray line is the high-frequency projection from interferometers. Two very important facts appear in this figure; First, the spectrum of the hypothetical GCE pulsars are depicted in black/red for the optimistic/pessimistic scenarios described in~\cite{Bartel:2024jjj}. Second, the BBN bound appears as the gray dotted contour. Cosmological sources of GWBs can have strains as high as this contour. As the blue contours show, the proposed experiment is able to detect the GCE pulsars, and to test some of the cosmological sources, as it is the case of 1st-order phase transitions marked by the green shadow.\\

\begin{figure}[t]
\centering
\includegraphics[width=0.48\textwidth]{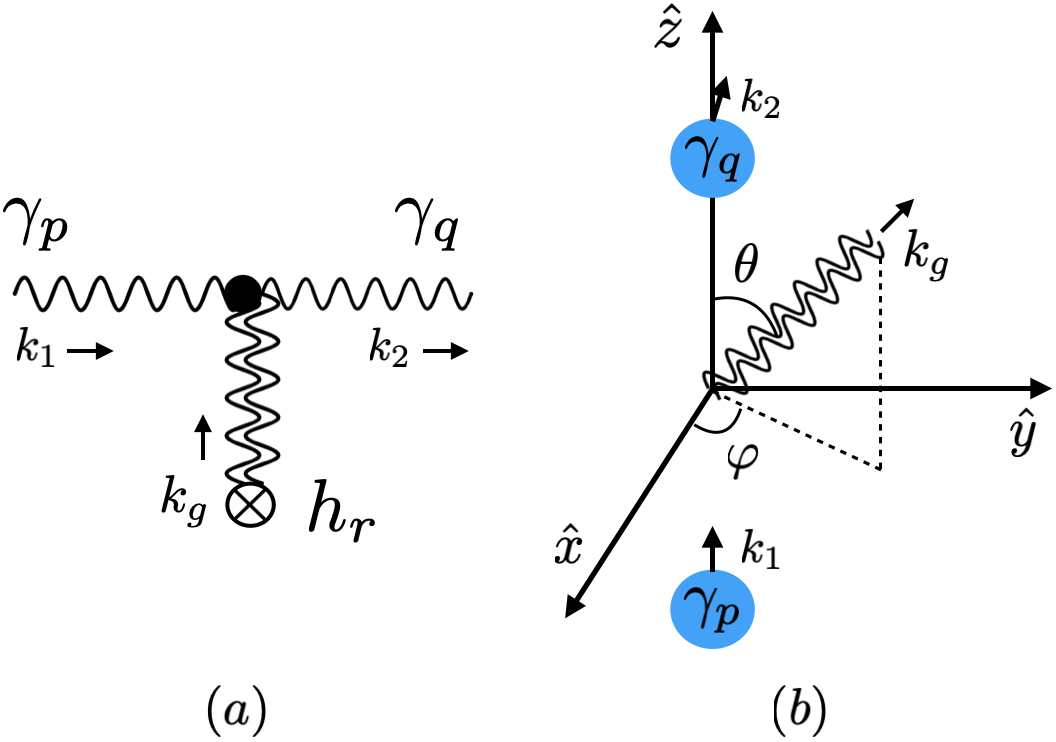}
\caption{(a) Feynman diagram of the interaction for an incoming photon with polarization $p$ and momentum $k_1$, a background gravitational wave with polarization $r$ and momentum $k_g$, and an outgoing photon with polarization $q$ and momentum $k_2$. (b) Kinematic configuration considered in this letter.}
\label{fig:diagram}
\end{figure}

{\bf Calculation} - The starting point to derive Eq.~(\ref{eq:Ns}) is to consider the free photon action
\begin{equation}
S = \int d^4x\sqrt{-g}\left[-\frac{1}{4}g^{\mu\nu}g^{\alpha\beta}F_{\mu\alpha}F_{\nu\beta}\right], \label{eq:S}
\end{equation}
and to expand the metric up to a linear perturbation $h^{\mu\nu}$ on the flat metric $\eta^{\mu\nu}={\rm diag}(-1,1,1,1)$. This means expanding $g^{\mu\nu}=\eta^{\mu\nu}-h^{\mu\nu}$ and $g={\det}(g_{\mu\nu})\approx1+h^{\alpha\beta}\eta_{\alpha\beta}/2$ and collecting terms linear in $h^{\mu\nu}$
\begin{eqnarray}
S &\supset& \int d^4x\,\,\, \frac{1}{2}h^{\alpha\beta}P_{\mu\nu\lambda\alpha\sigma\beta}F^{\mu\lambda}F^{\nu\sigma}, \label{eq:Ssub} \\
P_{\mu\nu\lambda\alpha\beta\sigma}&=&(\eta_{\mu\nu}\eta_{\lambda\alpha}\eta_{\sigma\beta} - \eta_{\lambda\sigma}\eta_{\mu\nu}\eta_{\alpha\beta}/4),\nonumber
\end{eqnarray}
where the electromagnetic field tensor is defined as usual $F_{\mu\nu}=\partial_\mu A_\nu-\partial_\nu A_\mu$.

At the particle level, the process of interest is depicted in Fig.~\ref{fig:diagram}(a). An incoming photon with momentum $k_1^\mu=\omega_1(1,\hat {\bf k}_1)\equiv \omega_1\hat k_1^\mu$ scatters off a GW with momentum $k_g^\mu=\omega_g(1,\hat {\bf k}_g)$ producing a photon with momentum $k_2^\mu=\omega_2(1,\hat {\bf k}_2)\equiv \omega_2\hat k_2^\mu$. The initial state photons are confined in a laser pulse, hence their wave functions are not plane waves spread through all space. To capture this effect one can consider the initial state to be a wave packet describing the field of the pulse. For a Gaussian wave packet of amplitude $E_0$, polarized along $\hat{\bf x}$, and traveling along $\hat{\bf z}$, the electric field is (the real part of)
\begin{eqnarray}
{\bf E}({\bf x},t)&=&E_0f(z,t)e^{i\omega_1(t-z)}\hat{\bf x},\\
f(z,t)&=&e^{-(t-z)^2/2\tau^2}.
\end{eqnarray}

The amplitude for the process of interest is given by
\begin{equation}
\mathcal{A}_{1\to2}=\left\langle \gamma_{k_2} \right| \frac{i}{2} \int d^4x\, h^{\alpha\beta}P_{\mu\nu\lambda\alpha\beta\sigma}F^{\mu\lambda}F^{\nu\sigma} \left| 0 \right\rangle.
\label{eq:amplitude}
\end{equation}

The wave function of the outgoing photon with polarization $p$, momentum $k$, and quantization volume $V$ is
\begin{equation}
A^{\lambda}(x)\left|\gamma_{k_2}\right\rangle =\frac{1}{\sqrt{2V\omega_2}}\epsilon^{\lambda}_p(k)e^{ik_2\cdot x}\left|0\right\rangle,
\label{eq:photon}
\end{equation}
and $h$ can be factorized the following way
\begin{equation}
h^{\alpha\beta} = h_0\epsilon^{\alpha\beta}_re^{ik_g\cdot x}.
\label{eq:h}
\end{equation}

The tensor $\epsilon^{\alpha\beta}_r$ is the polarization state $r$ of the GW. For a GW traveling along $\hat{\bf k}_g$, as in Fig.~\ref{fig:diagram}(b), one can write the polarization states $+$ and $\times$ in the Transverse Traceless (TT) gauge as $\epsilon^{ij}_{\rm +} = (U_iU_j-V_iV_j)/\sqrt{2}$ and $\epsilon^{ij}_{\rm \times} = (U_iV_j+V_iU_j)/\sqrt{2}$, with unit vectors $V = (-s_{\varphi},c_{\varphi},0)$ and $U = V\times\hat{\bf k}_g$ (where $s_{\alpha}\equiv\sin\alpha$ and $c_{\alpha}\equiv\cos\alpha$).

With these ingredients, the amplitude reads
\begin{equation}
\mathcal{A}_{1\to2}= \frac{h_0E_0\omega_2}{\sqrt{2V\omega_2}}\mathcal{T}_{pqr}\int d^4xf(z,t)e^{ix\cdot q},
\label{eq:amplitude_2}
\end{equation}
where $q^\mu=(k_1+k_g-k_2)^\mu$. The quantity $\mathcal{T}_{pqr}$ takes care of the tensor products
\begin{eqnarray}
\mathcal{T}_{pqr}&=&\epsilon^{\alpha\beta}_r{\bar P}_{\mu\nu\alpha\lambda\beta\sigma} \hat k_2^\mu\epsilon_q^{*\lambda}{\hat F}_p^{\nu\sigma} ,\\
{\bar P}_{\mu\nu\alpha\lambda\beta\sigma}&=& \eta_{\mu\nu}\eta_{\alpha\lambda}\eta_{\beta\sigma}+\eta_{\alpha\nu}\eta_{\sigma\lambda}\eta_{\beta\mu}-\eta_{\alpha\beta}\eta_{\mu\nu}\eta_{\lambda\sigma}/2,\nonumber
\label{eq:T}
\end{eqnarray}
and ${\hat F}^{0i}=E^i$, ${\hat F}^{ij}=\epsilon^{ijk}B^k$ describe the fields of the laser (normalized to one).

This letter focuses in the limit $\omega_g\ll \omega_1$ where the outgoing photons are highly collinear with the laser. Up to factors of $\omega_g/\omega_1$, the polarization vectors of the final state photons are $\epsilon_\pm = (0,\mp1,-i,0)/\sqrt{2}$. In the limit of interest the only non-zero tensor products are
\begin{equation}
\mathcal{T}_{\hat{x}\pm+}=\mp{1\over 2} s^2_{\theta}\,.
\label{eq:T_product}
\end{equation}

To obtain the transition probability one needs to integrate over the phase space of the final state photon
\begin{equation}
P = \int \frac{d^3{\bf k}_2}{(2\pi)^3} V_2 |\mathcal{A}_{1\to2}|^2.
\label{eq:prob}
\end{equation}
Because the initial state has been defined as the {\it total} field of the laser, this probability gives the {\it total} number of signal photons, hence $P=N_s$. When inserting Eq.~(\ref{eq:amplitude_2}) in (\ref{eq:prob}), one can replace the field amplitude by $E_0^2=N_\gamma \omega_L/A\tau$, where $\omega_L=\omega_1$ is the frequency and $A$ the transverse area of the pulse. Finally, adding the probabilities of the two transitions $\hat x \to \pm$ one gets
\begin{equation}
N_s=\frac{N_\gamma h_0^2\omega_L^2s^4_{\theta}}{4\tau}\int_0^Ldz\left| \int_0^Tdtf(z,t)e^{it(1-c_{\theta})\omega_g} \right|^2.
\label{eq:Nsvariant}
\end{equation}

\begin{figure}[t]
\centering
\includegraphics[width=0.45\textwidth]{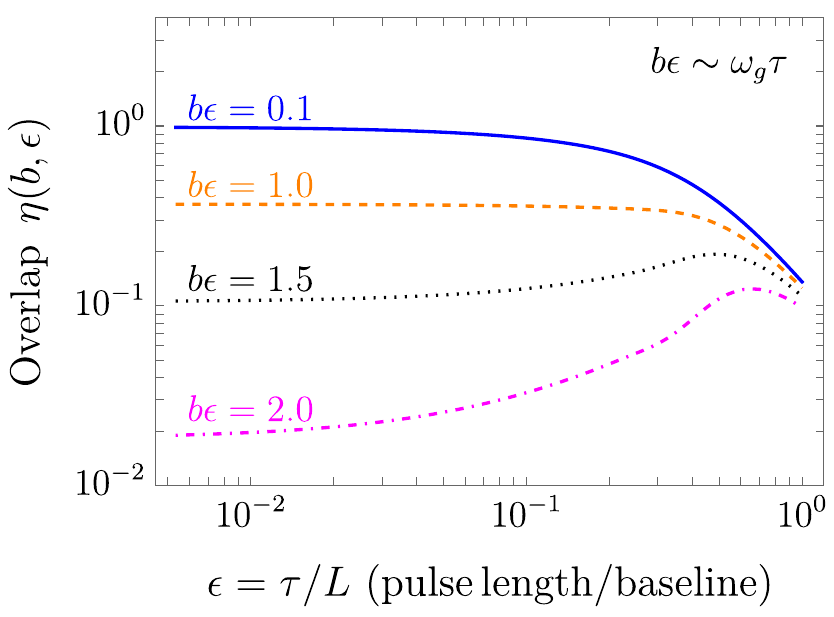}
\caption{Overlap coupling for the gravitational photon polarization twist effect defined in Eqs.~(\ref{eq:eta})-(\ref{eq:F0}).}
\label{fig:eta}
\end{figure}

This formula gives the number of signal photons after the pulse-GW overlap over a baseline $L$ and time $T$. One can rewrite Eq.~(\ref{eq:Nsvariant}) as Eq.~(\ref{eq:Ns}), which clearly shows how the effect increases with the baseline.

The {\it overlap coupling} $\eta$ in Eq.~(\ref{eq:Ns}) is defined by
\begin{eqnarray}
\eta(b,\epsilon)&=&\int_0^1d\alpha\left|F_0(\alpha,b,\epsilon)\right|^2,\label{eq:eta}\\
F_0(\alpha,b,\epsilon)&=&\frac{1}{\sqrt{    2\pi}}\frac{1}{\epsilon}\int_0^1d\beta e^{-(\alpha-\beta)^2/2\epsilon^2}e^{ib\beta}\label{eq:F0},
\end{eqnarray}
where $z\to\alpha L$, $t\to\beta L$, $\epsilon=\tau/L$, and $b=(1-c_{\theta})\omega_gL$. This coupling depends on two main parameters; $\epsilon$ which is a comparison between the pulse length and the baseline, and $b\epsilon=(1-c_{\theta})\omega_g\tau$ which is a comparison between the GW wavelength and the pulse length. Its behavior is shown in Fig.~\ref{fig:eta}. For a given pulse length and baseline (e.g., $\epsilon=10^{-2}$), the GW-pulse overlap is good as long as the wavelength of the GW is larger ($b\epsilon=0.1$, blue solid curve) or comparable to the pulse length ($b\epsilon=1.0$, orange dashed curve). When the GW wavelength becomes small compared to the pulse length ($b\epsilon>1.5$, dotted black and dash-dotted cyan curves) the effect is suppressed. This suggests the following technique for GW searches with this mechanism: First, match the laser pulse to be, at most, equal to the GW wavelength. Second, increase the baseline as much as possible. These choices maximize the effect. \\

{\bf Summary} - The idea that GWs can affect the polarization of photons is known. This letter harnesses this effect to investigate the detection of GWBs from the early Universe and the Galactic Center by means of a polarized laser pulse. The proposed experiment involves at least two satellites equipped with mirrors, a birefringent material, and a single photon detector. Based on conservative estimates, the findings suggest a promising approach. For the best results, the pulse length should match the wavelength of the targeted GW, providing a key test to confirm discovery: if a signal is detected, by increasing the pulse length, the strength of the signal should diminish. Additionally, the effect scales with the distance between the satellites; the longer the laser pulse travels through the GWB, the more signal photons are captured by the birefringent material.

Some of the assumptions made in this letter should be explored in more details in future studies. This includes studying non-linear effects within the high-power laser pulse and between the pulse and the CMB, as well as the evolution of the beam profile as it travels through the baseline. Background sources should also be studied in more details. For instance, cosmic rays could produce internal showers in the experimental setup which in turn could increase noise levels. As it was shown, a LISA-like laser can detect the hypothetical GCE pulsars if the laser pulse can be recycled bouncing between satellites to gain statistics. If recycling is not possible, a more powerful laser is needed (from W to kW). In such a case an important challenge is the technical feasibility of the laser pulse needed for the job and the practicality of deploying it up in space. Besides these, electronic noise, uncertainty in the laser polarization, and possible misalignment between the satellites and the polarization axis are among the technical issues that could emerge. Despite the foreseeable challenges in implementing the Fig.~\ref{fig:setup} setup, the physics case for this experiment is of utmost interest as it is shown in Fig.~\ref{fig:sensitivity}.

This letter introduces a new technique to measure gravitational wave backgrounds. These backgrounds could be from a diffuse population of astronomical sources or they could be cosmological i.e. the remnant of early Universe dynamics. As a concrete example, it is shown that this technique can detect the hypothetical pulsars responsible for the excess of gamma rays from the Galactic Center. The detection of this signal could settle a long debate within the high-energy physics community; a successful detection would confirm the presence of these pulsars, while a null result would lend further support to the dark matter hypothesis. Beyond the Galactic Center, the proposed experiment can probe cosmological sources like hypothetical 1st-order phase transitions in the early Universe. These sources appear in highly motivated theories beyond the Standard Model. The possibility of detecting cosmological GW signals allows the proposed experiment to shed light into theories about the origin of matter, reheating, inflation, etc., opening our eyes closer to the origin of time, the time of the creation, the time of the {\it Bang}. \\

{\bf Acknowledgments} - The author is thankful to Masaaki Sakakura for insightful comments on induced birefringence.
Special thanks to Tanner Trickle for pointing out an issue in the original calculation, and Asher Berlin and Alex Millar for suggestions on how to fix it. Also, to Kayla Bartel and Stefano Profumo for discussions on the GW spectrum of the GCE pulsars.
Further thanks for valuable feedback to Ryan Plestid, Tom Krokotsch, Federico Meloni, Nick Rodd and, especially, Henry Lamm.
This manuscript has been authored by Fermi Forward Discovery Group, LLC under Contract No. 89243024CSC000002 with the U.S. Department of Energy, Office of Science, Office of High Energy Physics.
%

\bibliography{ref}

\end{document}